\documentstyle[preprintaps,revtex]{aps}
\newcommand{\be}{\begin{equation}}
\newcommand{\ee}{\end{equation}}
\newcommand{\bea}{\begin{eqnarray}}
\newcommand{\eea}{\end{eqnarray}}
\newcommand{\half}{\mbox{$\frac{1}{2}$}}
\newcommand{\quart}{\mbox{$\frac{1}{4}$}}
\newcommand{\bm}[1]{\mbox{\boldmath$#1$}}
\newcommand{\intx}{\int\!d^3x\,}
\newcommand{\p}{\prime}
\math-with-secnums
\tightenlines
\begin{document}
\def\overlay#1#2{\setbox0=\hbox{#1}\setbox1=\hbox to \wd0{\hss #2\hss}#1%
\hskip -2\wd0\copy1}
\begin{title}
\begin{center}
Chiral colour-dielectric model with perturbative \\ quantum pions and gluons
\end{center}
\end{title}
\author{L. R. Dodd and D. E. Driscoll}
\begin{instit}
 Department of Physics and Mathematical Physics, University of Adelaide,\\
Adelaide, South Australia.
\end{instit}
\begin{abstract}

Pionic contributions to static nucleon properties are calculated in a
chiral extension of the colour-dielectric model. The pion field and residual
gluon field are treated perturbatively. It is shown that with a simple
choice for the energy of the scalar confining field and in the chiral limit,
the system of equations describing the bare soliton and the perturbative
pion and gluon fields may be cast in a dimensionless, parameter free form for
large glueball mass. This enables a  formula for the masses of the nucleon
and delta including leading order pionic and gluonic contributions and
corrections for spurious centre-of-mass motion, valid
for a wide range of input parameters determining the bare soliton solutions,
to be derived.
A further consequence of the scaling behaviour is that pionic contributions
to nucleon properties, calculated using the methods of the cloudy bag model,
are insensitive to the soliton parameters, once the size of the soliton is
fixed. The model results are very similar to those of the cloudy bag model
but the predicted masses are about 20\% too large, and the pionic contributions
to charge radii are underestimated.

\end{abstract}

\section{introduction}

In their simplest form  colour-dielectric models (CDM) \cite{reviews}
describe the quark structure of hadrons
by confining effective quark fields with a scalar field which represents the
long range order of the QCD vacuum. Like the MIT bag \cite{MIT} lagrangian,
 the typical lagrangian of these models at this level
 is not chirally symmetric, but it is well known that by introducing an
suitable interaction with an elementary pion field, manifest chiral
symmetry may be restored\cite{Thomas}. There is no unique prescription for the
additional terms in the effective lagrangian and a number of different
chiral versions of nontopological soliton models have been considered
by various authors (reviewed recently by Birse \cite{reviews}).

Following the approach which was used to obtain the cloudy bag model
(CBM) lagrangian \cite{Thomas} from the MIT bag model, Williams and Dodd
\cite{WD} investigated chiral extensions of both the Nielsen-Patkos
colour-dielectric model \cite{NP} and the Friedberg-Lee soliton bag
model \cite{FL}. It was found
that the pion fields in the soliton were sufficiently weak that pionic
contributions to nucleon properties could be calculated using
perturbation theory as in the CBM work \cite{Thomas,CBM,CJP,DTA}.
                                         This is to be contrasted with
non-perturbative approaches where the pion field is treated in the mean
field approximation using the hedgehog ansatz \cite{hedgehog}.
The numerical results of reference \onlinecite{WD} for pionic corrections
showed an insensitivity to the details of the unperturbed soliton solutions
and, when the scale of the soliton
solution was fixed to reproduce the proton charge radius, broad agreement
with the results of the CBM. However, no attempt
was made in this work to choose a parameter set which would also fit the
nucleon and delta masses when centre-of-mass  corrections and gluonic
corrections, discussed below, were included.

Another refinement of the CDM, necessary for the calculation of mass splittings
of the hadrons, is the retention of residual colour fields left over from
the coarse-graining of the QCD fields. For example the mass degeneracy of the
nucleon and delta isobars is lifted by the colour magnetic hyperfine
interaction. The one-gluon exchange contribution
to the nucleon-delta mass difference has been calculated in the CDM both
perturbatively and self-consistently \cite{gluons,DW}.
However, these calculations did not
take into account the contribution from pion exchange expected from the chiral
models.

The aim of the present work is to test the predictions of the CDM for static
nucleon properties including both pionic and gluonic contributions and with
centre of mass corrections. A similar calculation has appeared recently.
Leech and Birse \cite{Leech} have calculated pionic contributions using
Peierls-Yoccoz projection to remove spurious centre-of-mass contributions.
They use a chiral version of the CDM where the pion fields are accompanied
by an additional scalar field, as in the linear sigma model, rather than the
non-linear realization of chiral symmetry adopted in this paper. Although
their lagrangian has chiral symmetry, the Goldberger-Treiman relation,
which should be satisfied by the model, is violated by the approximations
made in projecting momentum eigenstates. In our work we have chosen to preserve
the Goldberger-Treiman relation at the expense of using only  cruder estimates
of centre-of-mass  corrections. In our view reliable estimates of c.m.
corrections
which respect the symmetries of the lagrangian remain a problem for these
models. Leech and Birse did not calculate the gluonic contribution to
the nucleon-delta mass splitting but assumed that the strength
of the quark-gluon coupling could be adjusted so that a fit to the non-pionic
part of the mass-splitting would be achieved. Here we calculate the M1
colour magnetic energy  explicitly to see whether consistent values of the
strong coupling constant are obtained over a range of soliton parameters.

We would like to emphasize that our model is just one of many possibilities.
{}From a more fundamental point of view it is natural to regard the pion (and
other mesons) as composites of the quark and gluon fields. For example
in the work of Banerjee {\it et al.} \cite{Banerjee} it is assumed that an
effective
low energy chiral model can be derived from QCD by entirely  eliminating
the gluon degrees of freedom in favour of meson exchanges between quarks.
In this approach one gluon exchange should not be added to the quark-meson
model. The Lagrangian that we use, as in the CBM, includes an additional
elementary pion field to restore chiral symmetry, and within the context
of the model both one pion and one gluon exchange are calculated.

Section 2 describes the chiral version of the colour-dielectric model
considered
in this paper, how the lowest order perturbative pionic and gluonic
contributions to the soliton energy are calculated, and how the masses of the
nucleon and delta are estimated including  c.m. corrections. The bare soliton
solutions are characterized by three parameters, the quark mass $m$, the
glueball mass $M_{\chi}$, and the scale $\sigma_v$ of the confining scalar
field. The magnitude of the gluonic energy shift is determined by the strong
coupling constant $\alpha_s$ which is essentally a free parameter of the
model. The magnitude of the pionic contributions are fixed through the
Goldberger-Treiman relation of the model in terms of the pion mass, the pion
decay constant and the axial coupling constant. The latter is calculated
from the bare soliton solution while the pion mass and pion decay constant
are given their experimental values. Thus once the bare soliton solutions
are chosen there is no further freedom in the model to vary the pionic
contributions to nucleon properties.

In section 3, following the scaling argument of McGovern, Birse and Spanos
\cite{MBS} for large glueball mass,  we are able to show
 that the system of equations determining the bare soliton solution and
the perturbative pion and gluon fields may be cast in a dimensionless,
parameter free form in the chiral limit where the pion is massless.
 This enables a mass formula for the nucleon and
delta masses to be given whose numerical coefficients are determined by
solving the universal equations once only. This scaling which still holds to a
good approximation for quite small ratios of the glueball to quark masses and
for non-vanishing pion mass explains the insensitivity of pionic corrections
to the soliton parameters found in earlier work \cite{WD}.

Pionic contributions to static nucleon properties are considered in
section 4. The formulae for charge radii and magnetic moments are essentially
identical with those of the cloudy bag model, with the CBM form factor
replaced by the form factor computed from the soliton solution.

Section 5 contains our numerical results and conclusions.

\section{the model}
\subsection{The Hamiltonian}
With the notation of reference \onlinecite{WD}
, the Hamiltonian of the chiral extension of
the colour dielectric model, including gluons, to be considered here may be
written as
\begin{equation}
H=H_{\rm{NS}}+H_{\pi}+H^{\pi}_{\rm{I}}+H_{g}+H_{\rm{I}}^{g}
\equiv H_{0}+H_{\rm{I}}^{\pi}+H_{\rm{I}}^{g}
\end{equation}
where the Hamiltonian for the non-topological soliton in the mean field
approximation (MFA) is
\begin{equation}
H_{\rm{NS}}=\intx \{:\bar{q}(i\mbox{\boldmath$\gamma . \nabla$}+m/\chi)q:
+\half \sigma_{v}^2(\bm{\nabla}\chi)^2 +\half \sigma_{v}^2 M_{\chi}^2 \chi^2\},
\end{equation}
the pion field contribution is
\begin{equation}
H_{\pi}=\intx\half\, :[(\partial_{0}\bm{\pi})^2+(\bm{\nabla}\bm{\pi})^2
+m_{\pi}^2 \bm{\pi}^2]:
\end{equation}
and the interaction between quarks and pions is given by
\begin{equation}
H_{\rm{I}}^{\pi}=\frac{i}{f_{\pi}} \intx \frac{m}{\chi}\,:\bar{q}\bm{\tau}.\bm
{\pi}\gamma_5 q:.
\end{equation}
The remaining terms in Eq. (2.1)
\begin{equation}
H_{g}= \intx (\kappa(\chi) F^{0\nu a}\partial_{0}A^{a}_{\nu}
+\quart \kappa(\chi) F^{a}_{\mu\nu}F^{\mu\nu a})
\end{equation}
and
\begin{equation}
H_{\rm{I}}^{g}=\half g_{s}\intx:\bar{q} \gamma^{\mu}\lambda^{a}A^a_{\mu}q:
\end{equation}
describe the coupling of effective gluon fields $A^{a}_{\mu}$ to the
colour singlet dielectric mean field $\chi$ through the dielectric function
$\kappa(\chi)=\chi^4$, and the quark fields $q$ respectively. As we consider
only single gluon exchange between quarks,
 the quadratic terms in the gluon field tensor
\begin{equation}
F^{a}_{\mu\nu}=\partial_{\mu}A^{a}_{\nu}-\partial_{\nu}A^{a}_{\mu}
+g_{s}f^{abc}A^{b}_{\mu}A^{c}_{\nu}
\end{equation}
are dropped, so that each of the gluon fields propagates like an independent
electromagnetic field in the presence of a spatially varying dielectric
medium. It should be noted that in the absence of a rigorous derivation
of the dielectric model from QCD, there is some arbitrariness in the
details of the Hamiltonian density adopted above. Bayer {\it et al.}
\cite{Bayer} and Banerjee \cite{Ban88} have argued that the quark-pion
coupling of Eq. (2.4) should be proportional to $\chi^{-2}$. The
 question of whether residual gluon interactions, Eq. (2.6), should be included
at all, has been mentioned in the introduction. However,  the work of
McGovern \cite{McG} in fitting the baryon spectrum with a chiral
dielectric model
including perturbative gluons lends some support to the model chosen here.
The fit using the inverse coupling of Eq.(2.4) was found to be more
satisfactory
than the fit using inverse square coupling.

In zeroth order the interactions between quarks and pions and
quarks and gluons may be ignored and the bare baryon states
are eigenstates of $H_{0}$ with no gluons or pions present. The bare nucleon
and delta states
 are thus described by the usual MFA solutions where the
mean $\chi$ field has spherical symmetry and the three quarks are all
placed in the lowest $1S$ mode. The upper and lower radial components
$u$ and $v$ of the quark wavefunctions and the quark energy eigenvalue
$\epsilon$ satisfy
\begin{eqnarray}
\frac{du}{dr}=-(\epsilon+\frac{m}{\chi})v,\\
\frac{dv}{dr}=(\epsilon-\frac{m}{\chi})-\frac{2v}{r}
\end{eqnarray}
and the mean field $\chi$ is determined self-consistently from
\begin{equation}
\frac{d^2\chi}{dr^2}+\frac{2}{\chi}\frac{d\chi}{dr}=
-\frac{3m}{\sigma_v^2\chi^2}(u^2-v^2)+M_{\chi}^2\chi
\end{equation}
with appropriate boundary conditions. The spin and isospin states
of the bare nucleon and delta, denoted here simply by $|A_{0}\rangle$,
are degenerate with energy
\begin{equation}
E_{0}=3\epsilon+2\pi\sigma_v^2\int_0^\infty\!dr\,r^2\left[(\frac{d\chi}{dr})^2
+M_{\chi}^2 \chi^2\right].
\end{equation}
\subsection{Perturbation theory}
Our aim is to include perturbative corrections due to one-pion and one-gluon
exchange. In the remainder of this
section we consider the mass splitting of the nucleon and delta to order
 $(1/f_{\pi})^2$ and $g_s^2$. In Sec. 4 pionic corrections to the static
nucleon properties will be evaluated.

Working in the Shr\"{o}dinger picture, we may write an exact formal equation
for the dressed nucleon or delta state $|A\rangle$ which satisfies
$H|A\rangle=E_{A}|A\rangle$,
\begin{equation}
|A\rangle=(Z_{2}^{A})^{\half}|A_{0}\rangle +(E_{A}-H_{0})^{-1}\Lambda
H_{\rm{I}}|A\rangle
\end{equation}
where both $|A\rangle$ and $|A_{0}\rangle$ are normalized to unity and
$\Lambda$ is the complement of the projection operator onto the space of
degenerate bare nucleon and delta states
\begin{equation}
\Lambda=\rm{I}-\sum_{A_0}|A_{0}\rangle\langle A_{0}| .
\end{equation}
The perturbation $H_{\rm{I}}=H_{\rm{I}}^g + H_{\rm{I}}^\pi$ includes
interactions with gluons
as well as pions. The energy shift $\Delta_{A}=E_{A}-E_{0}$ is determined from
\begin{equation}
\Delta_A=\langle A_0|H_{\rm{I}}|A_0\rangle
+\langle A_0|H_{\rm{I}}(E_0-H_0+\Delta_A)^{-1}\Lambda H_{\rm{I}}|A\rangle (Z_2
^A)^{-\half}
\end{equation}
The second order shift
\bea
\Delta_A^{(2)}&=&\langle A_0|H_{\rm{I}}(E_0-H_0)^{-1}\Lambda H_{\rm{I}}
|A_0\rangle\\
&=&\Delta_A^g+\Delta_A^{\pi}
\eea
is obtained by replacing $|A\rangle (Z_2^A)^{-\half}$ by $|A_0\rangle$ in
Eq. (2.14), noting that in this case $\langle A_0|H_{\rm{I}}|A_0\rangle$
vanishes and that the shift separates into distinct gluon and pion pieces.

\subsection{The pion shift}
A calculation of the pion shift, similar to that of Chin\cite{Chin}
for the MIT bag, yields
\be
\Delta^\pi_A = - \frac{8\pi}{3}\frac{m^2}{f_\pi^2}\sum_{i,j}\langle\bm{\sigma
_i}.\bm{\sigma_j}\,\bm{\tau_i}.\bm{\tau_j}\rangle_A\,\mbox{$\rm{M_\pi}$}
\ee
with
\be
\mbox{$\rm{M_\pi}$}=\int_0^\infty\int_0^\infty \frac{u(r)v(r)}{\chi(r)}
\Delta(r,r')\frac{u(r')v(r')}{\chi(r')}\,r^2dr\,r'^2dr'\,
\ee
where $\Delta(r,r')$ is the free pion propagator. (In his work Chin uses a pion
propagator which excludes the pion from the bag.) The question arises whether
the quark-pion self-energies given by the terms with $i=j$ should be included
in the sums over the spin-isospin matrix elements in Eq.(2.17). Chin excludes
the self-energies from the energy shift, grouping them with the vacuum energy
of
the bag. On the other hand in  cloudy bag model calculations, they are included
in order that intermediate quark states may be coupled together to give the
full subspace of intermediate nucleon and delta states. If $S$ is the spin and
$T$  the isospin of the state $A$, then\cite{Aerts}
\be
\sum_{i\neq j}\langle\bm{\sigma_i}.\bm{\sigma_j}\,
\bm{\tau_i}.\bm{\tau_j}\rangle_A=36-4S(S+1)-4T(T+1)
\ee
and $\sum_{i=j} \langle\ldots\rangle_A=27$ for both nucleon and delta.
Thus the predicted splitting of the energy levels of the nucleon and delta
due to the pion, to the order of approximation
considered here, does not depend on the pionic self-energies.

It is convenient to define
\be
\Pi(r)=\int_0^\infty\Delta(r,r')\frac{u(r')v(r')}{\chi(r')}\,r'^2dr',
\ee
satisfying
\be
\frac{d^2\Pi}{dr^2}+\frac{2}{r}\frac{d\Pi}{dr}-\frac{2\Pi}{r^2}
-m_\pi^2\Pi=\frac{uv}{\chi},
\ee
in terms of which
\be
\mbox{$\rm{M}_{\pi}$}=\int_0^\infty((\frac{d\Pi}{dr})^2
+\frac{2\Pi^2}{r^2}+m_{\pi}^2\Pi^2)\,r^2dr.
\ee

\subsection{The gluon shift}
The shift due to exchange of gluons in the
dominant $M1$ mode is
\be
\Delta_A^g=-\frac{4}{3}\pi g_s^2\sum_{i,j}\langle\bm{\lambda_i}.
\bm{\lambda_j}\,\bm{\sigma_i}.\bm{\sigma_j}\rangle_A\, \rm{M}_g
\ee
with
\be
\mbox{$\rm{M}_g$}=\int_0^\infty \int_0^\infty \frac{u(r)v(r)}{r\kappa(r)}
g(r,r') \frac{u(r')v(r')}{r'\kappa(r')}\,r^2 dr\,r'^2 dr'
\ee
where $g(r,r')$ is the static Green's function\cite{Bicke}
for the propagation of the confined $M1$ gluon.

The matrix elements of the quark spin and colour observables in Eq. (2.23)
are taken with respect to the spin-isospin-colour states of the nucleon
or delta. In the sum over quarks it is customary to exclude the terms with
$i=j$ i.e. colour magnetic self-energies of the quarks are not included,
and $\sum_{i\neq j}\langle\bm{\lambda_i}.\bm{\lambda_j}\,\bm{\sigma_i}
.\bm{\sigma_j}\rangle_A=\pm16$, the plus sign for the nucleon and the minus
for the delta.
This choice is supported by the derivation\cite{CKY}
of the shift using relativistic,
many-body perturbation theory which suggests that the quark self-energies
should be regarded as part of the vacuum energy of the soliton.
However, in the present work we ignore the Dirac sea and make no attempt
to calculate the Casimir energy of the soliton. As usual we assume that
the colour electric energies for  quarks in the same spatial state
sum to zero.

An equivalent expression\cite{DW} for $\rm{M}_g$ which avoids the
construction of the Green's function,
\be
\mbox{$\rm{M}_g$}=\int_0^\infty ((\frac{dF}{dr})^2
+\frac{2F^2}{r^2})\kappa\, dr,
\ee
uses the field function $F(r)$ which satisfies
\be
\frac{d^2F}{dr^2}+\frac{1}{\kappa}\frac{d\kappa}{dr}\frac{dF}{dr}-
\frac{2F}{r^2}=\frac{uvr}{\kappa}.
\ee
Eqs. (2.24) and (2.25) may be shown to be equivalent by using the explicit
expression for the Green's function and integration by parts.

\subsection{Center-of-mass corrections}
The nucleon and delta energies
\be
E_A=E_0+\Delta_A^g+\Delta_A^\pi
\ee
contain contributions  from the center-of-mass motion of the soliton.
Our calculated masses
\be
M_A=(E_A^2-\langle P^2\rangle_{A,q}-\langle P^2 \rangle_{A,\chi})^
{\half}
\ee
include corrections for the quark momentum\cite{Det}
\be
\langle P^2 \rangle_{A,q}=12\pi\int_0^\infty dr\,[r^2(\epsilon+
\frac{m}{\chi}v)^2+(-2v+r(\epsilon-\frac{m}{\chi})u)^2+2v^2]
\ee
and the momentum of the $\chi$ field\cite{Lub} (using a quantum
 coherent state to produce the mean $\chi$ field),
\be
\langle P^2 \rangle_{A,\chi}=2\pi M_{\chi}\sigma_v^2\int_0^\infty (\frac{d\chi}
{dr})^2r^2\,dr.
\ee

In our numerical calculations the differential equations (2.8) and (2.9) for
the quark wavefunctions, Eq. (2.10) for the $\chi$ field, Eq.(2.22) for the
pion
field and Eq.(2.26) for the gluon field,
together with a normalization integral for the quark wavefunctions,
 are formulated as a non-linear boundary
value problem and solved simultaneously.\cite{DL} In the next section we
demonstrate that this system has interesting scaling properties
leading to a formula for the delta and nucleon masses in the chiral limit
$m_\pi=0$.

\section{the nucleon and delta masses in the chiral limit}

\subsection{Scaling}
The MFA solutions describing the degenerate bare nucleon and delta states
depend on three parameters, the quark mass $m$, the glueball mass $M_{\chi}$
and
the scale $\sigma_v$ of the $\chi$ field. McGovern, Birse and Spanos\cite{MBS}
have shown that for sufficiently large values of the glueball mass only two of
the parameters are independent and after choosing one to fix the size of the
soliton, one is left with a one parameter family of MFA solutions. In this
section we extend their arguments to find a mass formula for the nucleon
and delta which includes the colour-magnetic energy, the pion interaction
energy
and corrections for centre-of-mass motion.

With the help of a length unit
\be
 r_0=(mM_{\chi}\sigma_v)^{-\frac{1}{3}},
\ee
new dimensionless variables may be introduced:
\bea
r&=& r_0 x,\\
\epsilon&=&r_0^{-1}\epsilon_0,\\
\chi&=&mr_0 \chi_0,\\
u&=&r_0^{-\frac{3}{2}} u_0,\\
v&=&r_0^{-\frac{3}{2}} v_0,\\
F&=&m^{-4}r_0^{-4}F_0
\eea
and
\be
\Pi=m^{-1}r_0^{-2}\Pi_0.
\ee
In terms of these variables, the system to be solved is
(the prime denotes differentiation with respect to $x=r/r_0$)
\bea
u_0^\p&=&-(\frac{1}{\chi_0} +\epsilon_0)v_0,\\
v_0^\p+\frac{2}{x}v_0&=&-(\frac{1}{\chi_0}-\epsilon_0)u_0, \\
(\chi_0^{\p\p}+\frac{2}{x}\chi_0^\p)\frac{1}{M_{\chi}^2 r_0^2}&=&\chi_0
-\frac{3}{\chi_0^2}(u_0^2-v_0^2),\\
F_0^{\p\p}+\frac{4\chi_0^\p}{\chi_0}F_0^\p-\frac{2}{x^2}F_0&=&\frac{u_0v_0x}
{\chi_0^4},\\
\Pi_0^{\p\p}+\frac{2}{x}\Pi_0^\p-\frac{2}{x^2}\Pi_0-m_{\pi}^2 r_0^2\Pi_0&=&
\frac{u_0 v_0}{\chi_0}
\eea
with the normalization condition
\be
4 \pi\int_0^\infty(u_0^2+v_0^2)x^2\,dx=1.
\ee
For sufficiently smooth variations of the $\chi_0$ field and large values
of the glueball mass $M_\chi$, the left hand side of Eq. (41) is negligible
and $\chi_0$ is simply determined from the quark wavefunction,
\be
\chi_0^3=3(u_0^2-v_0^2).
\ee
If furthermore the pion mass vanishes, Eqs. (3.9), (3.10), (3.12)-(3.15)
constitute a {\it dimensionless,  parameter free system} which need be solved
only once to determine the quark wavefunctions, the $\chi$ field, and the pion
and gluon fields for all values of $m$, $\sigma_v$ and $M_{\chi},$ provided
$M_{\chi}r_0$ is large.

\subsection{Mass formula}
Evaluation of the energy of the nucleon in terms of the scaled variables gives
\be
E_N=r_0^{-1}[3\epsilon_0+c_1+c_2(M_{\chi}r_0)^{-2}+c_3 g_s^2(mr_0)^{-4}
+c_4(f_\pi r_0)^{-2}],
\ee
where $\epsilon_0$, $c_1$, $c_2$, $c_3$ and $c_4$ are
the constants:
\bea
\epsilon_0&=&2.426\\
c_1&=&2\pi\int \chi_0^2 x^2\,dx=1.456,\\
c_2&=&2\pi\int (\chi_0^\p)^2\,dx=6.68,\\
c_3&=&-\frac{256}{3}\pi^2\int[(F_0^\p)^2+\frac{2}{x^2}F_0^2]\chi_0^4\,dx
=-0.04617
\eea
and
\be
c_4=-80\pi\int[(\Pi_0^\p)^2+\frac{2}{x^2}\Pi_0^2]x^2\,dx=-0.1469,
\ee
determined from  numerical solution of the soliton equations in the limit
where Eq. (3.15) is satisfied. A typical solution is shown in Figure 1.
In the expression for the energy, Eq. (3.16), the first term is the quark
energy,
the second the potential energy of the $\chi$ field, the third the kinetic
energy of $\chi$, the fourth the colour-magnetic energy, and the fifth the
pion field energy (in the chiral limit). The energy of the delta is also given
by Eq. (3.16) with $c_3$ replaced by $-c_3$ and $c_4$ replaced by $c_4/5$.

The colour-magnetic energy appears to be strongly dependent on the quark mass.
However, as  McGovern  \cite{McG} points out,
the definition of the strong coupling constant $\alpha_s$ and the
dielectric function are inter-dependent and there is no unique value of
the quark-gluon coupling.
 From Eqs.(2.5), (2.6) and (2.7) we see
that a change
 $\kappa\rightarrow\lambda^4\kappa$ is compensated by the changes
 $A^a_{\mu}\rightarrow\lambda^{-2}A^a_{\mu}$ and $g_s\rightarrow\lambda^2g_s$.
We will fix the definition of the strong coupling constant by choosing
$\kappa=1$ at the center of the soliton. Since the value of the dielectric
function is
proportional to
 $\chi^4(0)=(mr_0)^4\chi^4_0(0)$ at the center of the soliton,
 the  coupling constant
 is $4\pi\alpha_s=g_s^2(mr_0\chi_0(0))^{-4}$ and the colour-magnetic
energy may be written as
\be
 4\pi\alpha_s c_3 \chi_0(0)^4 r_0^{-1}=0.9050 \alpha_s r_0^{-1},
\ee
which shows the expected dependence on the soliton parameters. Of course
defining the coupling constant and gluon potentials in this way does not
remove the sensitivity  of the colour magnetic energy to the
$\chi$ field inside the soliton. Once the scale of the $\chi$ field is set
variations of the field inside solitons with different quark content
will produce large relative changes in the gluonic energy.

The corrections, Eqs. (2.29) and (2.30), to the energy due to the
center-of-mass
motion also scale:
\be
\langle P^2 \rangle_q=c_5 r_0^{-2}
\ee
with
\bea
c_5&=&12\pi\int\{[(\epsilon_0+\chi_0^{-1})v_0]^2+[-2v_0/x+(\epsilon_0
-\chi_0^{-1})u_0]^2+2v_0^2/x^2\}x^2\,dx,\\
&=&16.12     \nonumber
\eea
and
\be
\langle P^2 \rangle_{\chi}= c_2(M_{\chi}r_0^3)^{-1}.
\ee

The nucleon and delta masses in the model are found from Eq. (2.28),
using Eq. (3.16) for the energies and Eqs. (3.23) and (3.24) for the momentum
corrections.

The parameter $r_0$ is related to the root mean square radius of the quark
distribution in the nucleon $R$ by
\be
R=\eta \langle x^2 \rangle^{\half} r_0
\ee
where
\be
\langle x^2 \rangle=4\pi \int(u_0^2+v_0^2)x^4\,dx=(0.7923)^2
\ee
and
\be
\eta=[1-2\lambda+3\lambda^2+\frac{3}{2}(\frac{\lambda}
{\epsilon_0 r_0})^2]^{\half}
\ee
with
\be
\lambda=\frac{\epsilon_0}{M_A r_0}\,,
\ee
is an additional scaling factor\cite{Det}
which estimates the reduction in size after removal of the motion of the
center-of-mass.

We note that the simple scaling behavior derived here depends on our initial
choice of the quadratic form of the potential energy of the $\chi$ field.
In practice this means that the MFA solutions are one-phase solutions in the
nomenclature of Ref. \onlinecite{DWT}.
 Unlike the usual bag models, there is no bag pressure,
 the energy of the $\chi$ field having the same $1/r_0$ dependence as the quark
energy. For two-phase solutions, possible in quartic
potentials, where there is rapid variation in $\chi$ between the interior
and exterior of the soliton, the kinetic energy of the $\chi$ field is not
negligible and the above scaling does not hold. Of course in this case
the full equations may be solved numerically for a given parameter set
which may include a bag pressure, but the simplicity of the energy formula
Eq. (3.16) is lost.

{}From the work of this section, we see that the masses of the nucleon and
delta
are essentially determined by the length scale $r_0=(mM_{\chi}\sigma_v)^
{-\frac{1}{3}}$ and the strong coupling constant $\alpha_s$, the pion decay
constant $f_{\pi}=93$ MeV being taken from experiment, and the small
corrections
due to the kinetic energy of the $\chi$ field being of order
$(M_{\chi}r_0)^{-1}
$. If $r_0$ is fixed by fitting the isoscalar charge radius of the nucleon and
$\alpha_s$ by fitting the nucleon-delta mass splitting, the predicted
 masses of the nuclon and delta in the model show little variation for a wide
range of quark and glueball masses.
In Table 1 the predictions for the masses  using the approximation Eq. (3.15)
are compared with those given by numerical solution of the full system,
Eqs. (3.9)-(3.14), for three parameter sets. The input parameters have been
fixed by reqiring a nucleon-delta mass splitting of 295 MeV and an isoscalar
nucleon radius of .75 fm in the full numerical calculations. Even for
ratios of the glueball mass to the quark mass as small as $M_{\chi}/m=10$
the approximate formula is remarkably accurate.

\section{Pionic corrections to nucleon properties}

Previous work\cite{WD} has shown that it is consistent to treat
the weak pion field in chiral non-topological solitons perturbatively, as is
done in the cloudy-bag model. With an appropriately  modified
pionic form factor, the CBM expressions may be applied to evaluate pionic
contributions to nucleon properties in  the present model.

The vertex functions for the absorption or emission of a pion are found by
expanding the pion field in a plane wave basis and taking matrix elements
of the interaction (2.4) between the bare soliton states $|A_{0}\rangle$.
In particular the vertex function $v^{AB}_{j}(\bm{k})$ for the absorption
of a pion with isospin $j$ and momentum $\bm{k}$ on the bare baryon state
$|B_{0}\rangle$ to produce the baryon state $|A_{0}\rangle$ may be written
as\cite{WD}
\be
v^{AB}_{j}(\bm{k})=-i\frac{f^{AB}}{m_{\pi}}\frac{\mu(k)}{(2\pi)^{3/2}(2\omega
_{k})^{1/2}}\sum_{m,n}\langle S_B,s_B,1,m|S_A,s_A\rangle
\langle T_B,t_B,1,n|T_A,t_B\rangle k_m^{*}e_{j,n}^*,
\ee
where $S_A$ and $s_A$ denote the spin and third component of spin for A
(  and similarly $T_A$ and $t_A$ for isospin), $k_m$ and $e_{j,n}$
are the spherical tensor components of the momentum $\bm{k}$ and the vector
$\bm{e}_{j}$ respectively, and $\omega_k^2=k^2+m_{\pi}^2$.
The CBM form factor \cite{CBM}
\be
\mu^{\prime}(kR)= 3j_{1}(kR)/kR ,
\ee
where $j_1$ is the spherical Bessel function of order one and $R$ is the bag
radius, is replaced in Eq. (4.1) by the  soliton form factor
\be
\mu(k)=\frac{\int drr^{3}(m/\chi (r))u(r)v(r)\mu^{\prime}(kr)}{\int drr^{3}
(m/\chi (r))u(r)v(r)},
\ee
defined so that $\mu(0)=1$. The form factors are compared in Figure 2.

 From Eqs. (1.8) and (1.9) it is easy to establish
\cite{WD,GDU}
that the denominator in (4.3) is proportional to the bare axial vector coupling
constant
\be
g_A^b=\frac{5}{3}\int d^{3}r(u^2(r)-\frac{1}{3}v^2(r))=\frac{20}{9}4\pi
\int drr^{3} \frac{m}{\chi}u(r)v(r)
\ee
and hence that the nucleon-nucleon transition coupling constant in (4.1) is
\be
f^{NN}=\frac{3}{2}\frac{m_\pi}{f_\pi}g_A
\ee
and that the other relevant couplings have the usual CBM ratios,
\be
f^{NN}:f^{\Delta \Delta}:f^{N\Delta}:f^{\Delta N}=
5:5:4\sqrt{2}:2\sqrt{2}.
\ee
In terms of the usual $\pi NN$ coupling constant,
$f^{NN}=(3 m_{\pi}/ 2 m_N)g_{\pi NN}=3\sqrt{4\pi}f_{\pi NN} $
and Eq. (4.5) is an expression of the Goldberger-Treiman relation.

\subsection{Scaling}

The solutions of Eqs.(2.8) and (2.9) for the quark wavefunctions and Eq.(2.10)
for the $\chi$ field may be used to construct the form factor (4.3) and the
transition coupling constants (4.6) using (4.4) and (4.5).
With the coupling constants and the form factor calculated from the soliton
solution replacing the CBM form factor and coupling constants, the usual CBM
expressions  for the pionic contributions to the nucleon and
delta self-energies, charge radii and magnetic moments
 etc. apply. Before discussing these contributions in detail, it is important
to note that the pionic corrections will be largely independent of the
 choice of the bare soliton parameters. This can be seen by applying the
transformations (3.2)-(3.6) of the previous section to the Eqs.(4.3) and (4.4).
The bare axial constant becomes
\be
g_A^b =\frac{80\pi}{9}\int dxx^3\frac{u_0 v_0}{\chi_0}.
\ee
For sufficiently large glueball mass, the scaled variables approach their
limiting forms, $g^b_A=1.318$, and hence  $f^{NN}$,$f^{\Delta \Delta}$,
$f^{N \Delta}$ and $f^{\Delta N}$ are constant under variation of  the soliton
parameters. In the same limit the form factor only depends on the length scale
set by $r_0=(M_{\chi}m\sigma_v)^{-\frac{1}{3}} $,since
\be
\mu(k)=\frac{80\pi}{9g_A^b}\int dxx^3\frac{u_0 v_0}{\chi_0}\mu^{\prime}(kr_0x).
\ee

\subsection{Pionic self-energies}
The pionic self-energies of the nucleon and delta are given by
\be
\Sigma^A=-\frac {1}{12\pi^2}\sum_{B} (\frac{f^{AB}}{m_{\pi}})^2
\int dk \frac{k^4\mu^2(k)}{\omega_k(\omega_k+m_B -m_A)}
\ee
In the cloudy bag model the masses $m_B=m_N,m_{\Delta}$ are usually taken as
the physical masses and renormalized perturbation theory is considered. Here,
since  we are only considering the leading order in a perturbative
calculation, the masses $m_N$ and $m_{\Delta}$ are equal to the bare soliton
mass. In this case the energy shift given by Eq. (4.9) is the same as that
given by Eq. (2.17), derived at the quark-pion level, provided the quark-pion
self-energies (terms with $i=j$ in Eq.(2.17)) are included. If the scaling
transformations
are applied to (4.9), we see that the mass splitting of the delta and nucleon
due to pions has the form $(f_{\pi}r_0)^{-2}r_0^{-1}I(m_{\pi}r_0)$ where $I$
is an integral depending on a single parameter, the product of the soliton
scale
and the chiral symmetry breaking pion mass.

\subsection{Electric form factors and charge radii}

The pionic contribution to the nucleon electric form factor is
\bea
G^{\pi}_{E,N}(q^2)&=&\pm\frac{1}{36\pi^3}(\frac{f^{NN}}{m_{\pi}})^2 \int d^3k
\frac{\mu(k)
 \mu (k') \bm{k} . \bm{k} '}{\omega_{k} \omega_{k'}(\omega_{k} +\omega_{k'})}
\\ & &
 \mp \frac{1}{72\pi^3}(\frac{f^{N \Delta}}{m_{\pi}})^2 \int d^3k
\frac{\mu (k)
\mu (k') \bm{k} . \bm{k} '}{(\omega_{k} +\omega_{\Delta N})(\omega_{k'}
+\omega_{\Delta N})(\omega_{k} +\omega_{k'})},   \nonumber
\eea
where $\bm{k}'=\bm{k}+\bm{q}$ and $\omega_{\Delta N}=m_{\Delta}-m_{N}$.
The upper sign holds for the proton and the lower for the neutron.
 Since (4.10) involves the difference of two similar terms, it turns out
that the calculated values of the electric root mean square radii of the
neutron and proton are quite sensitive to the assumed value of $\omega_{\Delta
 N}$. In our simple perturbative approach where $m_{\Delta}$ and $m_N$ are
equal to the bare soliton mass $\omega_{\Delta N}=0$. Alternatively, we may
compute the pionic correction, after the gluonic hyperfine splitting has been
calculated, by setting $\omega_{\Delta N}=\Delta_{\Delta}^g-\Delta_{N}^g$ (c.f.
 Eq. (2.23)). Numerical results for both choices are
compared in the next section.

The quark contribution to the electric form factor is proportional to the
Fourier transform of the quark density,
\be
G^{q}_{E,N}(q^2)=C_{N} \int d^3r(u^2(r)+v^2(r))e^{i\bm{q} \cdot \bm{r}},
\ee
where the constant $C_N$ is determined from charge conservation,
$G^{q}_{E,p}(0)+G^{\pi}_{E,p}(0)=1$ for the proton and $G^{q}_{E,n}(0)
+G^{\pi}_{E,n}(0)=0$ for the neutron.

The charge radii are calculated from the electric form factors by
\be
\langle r^2 \rangle_N =-6\frac{\partial}{\partial q^2}[G_{E,N}^{q}(q^2)+
G_{E,N}^{\pi}(q^2)]_{q^2=0}.
\ee
\subsection{Magnetic moments}
The pionic contribution to the nucleon magnetic moment is
\bea
\mu_{N}^{\pi}&=&\pm \frac{1}{27 \pi^2}(\frac{f^{NN}}{m_\pi})^2 \int_0^{\infty}
 \nonumber
dkk^4\frac{\mu^2 (k)}{\omega^4_k}\\& & \pm \frac{1}{216
\pi^2}(\frac{f^{N\Delta}
}{m_\pi})^2 \int_0^\infty dkk^4 \frac{\mu^2(k)(\omega_{\Delta N}+2\omega_k)}
{\omega_k^3(\omega_{\Delta N}+\omega_k)^2},
\eea
the upper sign holding for the proton and the lower for the neutron.

The quark contribution involves several integrals which determine the
probabilities of various components of the dressed nucleon. Define
\be
P_{BC\pi}=\frac{1}{12\pi^2}Z_{2}\frac{f^{NB}f^{NC}}{m_{\pi}^2}
\int_0^\infty dkk^4 \frac{\mu^2(k)}{\omega_k (\omega_{BN}+\omega_k)
(\omega_{CN}+\omega_k)},
\ee
then for the proton,
\be
\mu ^q_p=\frac{\mu_0}{27}(27 Z_2+P_{NN\pi}+20 P_{\Delta \Delta \pi}
+16\sqrt{2} P_{N\Delta\pi}),
\ee
and for the neutron,
\be
\mu ^q_n=-\frac{\mu_0}{27}(18 Z_2+4P_{NN\pi}+5P_{\Delta \Delta \pi}
+16\sqrt{2}P_{N\Delta \pi}).
\ee
The contribution from three bare quarks in Eqs.(4.15) and (4.16)  is
\be
\mu_0=\frac{2}{3} \int_0^{\infty}drr^3u(r)v(r),
\ee
and the normalization is determined by $Z_2+P_{NN\pi}+P_{\Delta
\Delta \pi}=1$.  The pionic and quark contributions together give
\be
\mu_N=\mu_N^q+\mu_N^{\pi}.
\ee

\section{Numerical Results and Conclusions}

Typical results of our numerical calculations of static nucleon properties,
 including pionic contributions, are shown in Table 2. In these calculations
chiral symmetry is broken by using the experimental value of the pion mass
in the field equation (3.13). Nevertheless Table 2 shows that the scaling
behaviour derived in section III for massless pions persists;
 when the overall scale of
the unperturbed soliton solution is set by matching the
experimental isoscalar charge radius, there is little variation in the
predicted nucleon properties over a wide
range of input parameters for the soliton.
As in other soliton bag calculations, the predicted nucleon masses are too
large. The nucleon mass can be made smaller by including the pion-quark
self-energies, but following the discussion of section II
, we believe it is more consistent not to do so.
The pion generates about
25\%  of the nucleon-delta mass splitting; the rest
is provided by the M1 colour magnetic splitting ($\alpha_s$ is adjusted to
reproduce the experimental mass difference). Comparison of Table 1 and
Table 2 shows that the pionic energy shift is decreased by about 17\% in going
from the massless to the massive pion.

The strength of the pion coupling is fixed by the experimental values of the
pion decay constant and the pion mass, and the value of the bare axial coupling
$g_A^b$, which is calculated from the model. The axial coupling is independent
of the soliton scale and is nearly constant over the various parameter sets.
Thus the pion coupling is almost constant and as can be seen from the first
three columns of Table 2, once the soliton size is
fixed, the pionic contributions to the nucleon
properties exhibit little dependence on the details of the soliton solutions.

The model is qualitatively similar to the Cloudy Bag Model; the pion cloud
increases the charge radius of the proton and gives the neutron a negative
charge radius. In Figure 3, we plot the neutron charge density, together with
its quark and pion components, for the soliton model using the parameters of
Column 2, Table 2 ($M_{\chi}$=1908 MeV). The quark and pion components for
the CBM are also shown for comparison, where the bag radius $R$= 1.1 fm.
This choice of $R$ minimises the root-mean-square difference of the soliton
and CBM form factors (shown in Figure 2). The quark (isoscalar) radius is then
20\% larger in the soliton model than in the CBM. The pionic term
$\langle r^2 \rangle _{\pi}$ is 10\% smaller. The decrease in
$\langle r^2 \rangle _{\pi}$ is attributable to the small tail of the form
factor (Figure 2).
 There are minor differences in the numerical values of the coupling
parameters   $f^{NN}$ chosen in the two models, some second order
renormalization being taken into account in the CBM, but these differences are
compensated by the larger value  for the mass difference
$\omega_{\Delta N}$ in the CBM where the physical masses are assumed.
Excluding centre-of-mass corrections, the proton and neutron charge radii are
.93 fm and -.31 fm in the soliton calculation in comparison with .87 fm and
-.34
fm in the corresponding CBM calculation \cite{CJP}. The values of $f_{NN\pi}$
and $g_A$ in Table 2 are bare values. They are decreased by about 5\% if the
renormalization procedure of the CBM is adopted.

 The values for the magnetic moments and charge radii given in Table 2 have
been
calculated with the difference between the delta and nucleon mass given by the
colour magnetic energy, $\omega_{\Delta N}= 2 \Delta^g_N$
in the energy denominators of Eqs. (4.10) and (4.13). For the parameter
set with glueball mass $M_{\chi}$=1908 MeV, the bracketed values, calculated
with a vanishing delta nucleon mass difference,
 are also given for comparison. It can be seen that the pionic contributions,
particularly the neutron charge radius, are quite sensitive to the assumed
value of $\omega_{\Delta N}$.

 Although the $\pi NN$ coupling constant agrees well with the
experimental value, pionic contribitions to static electromagnetic properties
are somewhat underestimated by the  model.
{}From the above comparison with the CBM it apears the pion form factor (4.3)
falls off too quickly in  momentum space compared with CBM form factor (4.2).
This can be compensated for by regarding the pion coupling $f^{NN}$
as an adjustable parameter rather than taking the value fixed by
Goldberger-Treiman relation, Eq.(4.5), of the model.
For example the results listed in the second last
column of Table 2, obtained with pion coupling strength increased by 40\%
are in excellent agreement with the experimental values.
However, it should be emphasized that this procedure violates one of the
attractive
theoretical features of the model, the connection between the scale of the bare
soliton and the magnitude of the pion cloud. In view of the
simplicity of the perturbative calculation of gluonic and pionic
corrections, and the crudeness of the estimates of centre-of-mass effects,
the quantitative results are in reasonable agreement with experiment.

\acknowledgments

We wish to thank A. G. Williams for helpful discussions and the
Australian Research Council for financial support.
One of us (L.R.D.) gratefully acknowledges the hospitality of the
Department of Physics and the Supercomputer Computations Research
Institute, Florida State University, while part of this work was carried
out.

\figure{Solutions of Eqs.(3.9)-(3.11) with a large value of the glueball
mass, $M_{\chi}r_0$=48.7. The dashed curve is the gluon source term of
Eq.(3.12).
For $x<1.4$ these curves are indistinguishable from the solutions of Eqs. (3.9)
and Eq. (3.10) with the approximation (3.15). \label{fig1}}
\figure{Comparison of the soliton model (solid line) and CBM (dashed line)
pion form factors for the parameter set of Column 2 of Table 2. The CBM radius,
R=1.1 fm, minimimises the root-mean-square difference of the form factors.
\label {fig2}}
\figure{Comparison of the quark and pion components of the neutron charge
density in the soliton model (solid curves) and the CBM (dashed curves). The
CBM radius is R=1.1 fm. The total charge density is shown only for the soliton
model.\label{fig3}}

\begin{table}
\caption{Nucleon and delta masses in the chiral limit. Quantities in
columns labelled (a) have been calculated by solving the full set of
soliton equations; the corresponding predictions from the approximate mass
formula are contained in columns (b).}
\begin{tabular}{lrrrrrr}
&\multicolumn{2}{c}{$M_{\chi}/m=10$} & \multicolumn{2}{c}{$M_{\chi}/m=50$}
&\multicolumn{2}{c}{$M_{\chi}/m=150$}\\
&\multicolumn{2}{c}{$M_{\chi}=648.6MeV$}&\multicolumn{2}{c}
{$M_{\chi}=1910.5MeV$}&\multicolumn{2}{c}{$M_{\chi}=3978.8MeV$}\\
&\multicolumn{2}{c}{$\alpha_s=0.6016$}&\multicolumn{2}{c}{$\alpha_s=0.5699$}
&\multicolumn{2}{c}{$\alpha_s=0.5687$}\\
\tableline
&(a)&(b)&(a)&(b)&(a)&(b)\\
\tableline
$\langle r^2 \rangle^{\half}(fm)$ & 0.750 & 0.752 & 0.750 & 0.751 & 0.750 &
0.750 \\
$E_0(Mev)$ & 1600.8 & 1624.5 & 1558.3 & 1559.2 & 1553.1 & 1553.2 \\
$\Delta_N^g(MeV)$ & $-$92.5 & $-$95.9 & $-$91.8 & $-$91.5 & $-$91.6 & $-$91.4
\\
$\Delta_N^{\pi}(MeV)$ & $-$97.9 & $-$92.7 & $-$95.3 & $-$94.8 & $-$95.3 &
$-$95.1\\
$M_N(MeV)$ & 1228.6 & 1227.1 & 1175.6 & 1165.6 & 1167.7 & 1162.1 \\
$M_{\Delta}(MeV)$ & 1523.6 & 1529.7 & 1470.6 & 1461.5 & 1462.7 & 1457.7

\end{tabular}
\end{table}

\begin{table}
\caption{  Results for nucleon properties, including perturbative pions and
 gluons, for four different soliton parameter sets. The soliton scale is
fixed to give the experimental isoscalar charge radius  and
the strong coupling constant to give the experimental value of the
nucleon-delta mass splitting in all cases. Centre-of-mass corrections are
included.}
\begin{tabular}{lrrlrrr}
$M_{\chi}/m$ & 10 & 50 & & 150 & 50 & \\
$M_{\chi}(MeV)$ & 648 & 1908 & & 3974 & 1920 &\\
$\alpha_s$ & 0.646 & 0.612 & & 0.611 & 0.401 & $Expt$\\
\tableline
$\langle r^2_0 \rangle^{\half}(fm)$ & 0.750  & 0.750 & & 0.750 & 0.750 & 0.750
\\
$E_0(Mev)$ & 1601  & 1556 & & 1551 & 1566 & \\
$\Delta_N^g(MeV)$ & $-$99.2 & $-$98.5 & & $-$98.2 & $-$65.0 & \\
$\Delta_N^{\pi}(MeV)$ & $-$81.6 & $-$79.5 & & $-$79.2 & $-$160.0 &\\
$M_N(MeV)$ & 1240 & 1185 & & 1177 & 1139 & 939 \\
$M_{\Delta}-M_{N}(MeV)$ & 295 & 295 & &295  &295 & 295 \\
\tableline
$f_{NN \pi}$ & 0.283 & 0.280 & & 0.279 & 0.393 & 0.28 \\
$g_{A}^{b}$ & 1.34 & 1.32 & & 1.32 & 1.32 & 1.27 \\
$\mu_p$ & 2.28 & 2.29 & (2.40)& 2.29 & 2.71 & 2.76 \\
$\mu_n $ & $-$1.72 & $-$1.72 & ($-$1.78) & $-$1.72 & $-$2.18 & $-$1.91 \\
$\langle r^2_p \rangle^{\half}(fm)$ & 0.795 & 0.795 & (0.770) & 0.795
& 0.830 & 0.83 \\
$\langle r^2_n \rangle^{\half}(fm)$ & $-$0.264 &  $-$0.263 & ($-$0.175)
& $-$0.262 & $-$0.355  & $-$0.35
\end{tabular}
\end{table}


\begin{references}
\bibitem{reviews} Extensive references are contained in the recent reviews
 by M. C. Birse, Prog. Part. Nucl. Phys. {\bf 25}, 1 (1990), and by
L. Wilets, {\it Nontopological Solitons},(World Scientific, Singapore
,1989).

\bibitem{MIT} A. Chodos {\it et al.},Phys. Rev. D{\bf 9}, 3471 (1974);
T. DeGrand {\ et al. , ibid.} {\bf 12},2060 (1975); C. E. DeTar and J. F.
Donoghue, Annu. Rev. Nucl. Part. Sci. {\bf 33}, 235 (1983); P. Hasenfratz
and J. Kuti, Phys. Rep. {\bf 40}C, 75 (1978).

\bibitem{Thomas} See, for example, the review by A. W. Thomas, in {\it Advances
in Nuclear
Physics}, edited by J. W. Negele and E. W. Vogt (Plenum, New York, 1983),
Vol. 13, p. 1.

\bibitem{WD}A. G. Williams and L. R. Dodd, Phys.\ Rev.\ {\bf D37}, 1971
 (1988).

\bibitem{NP} G. Chanfray, O. Nachtmann and H. J. Pirner, Phys. Lett.
{\bf 147B}, 249 (1984); H. B. Nielsen and A. Patkos, Nucl. Phys. {\bf B195}
, 137 (1982).

\bibitem{FL} R. Friedberg and T. D. Lee, Phys. Rev. D {\bf 15}, 1694 (1977);
T. D. Lee, {\it Particle Physics and Introduction to Field Theory} (Harwood
, Chur, Switzerland, 1981); R. Goldflam and L. Wilets, Phys. Rev. D {\bf 25},
 1951 (1982).

\bibitem{CBM} A. W. Thomas, S. Th\'{e}berge and G. A. Miller, Phys. Rev.
D {\bf 24}, 216 (1981); S. Th\'{e}berge and A. W. Thomas, Nucl. Phys.
{\bf A393}, 252 (1983); S. Th\'{e}berge, Ph. D. thesis, University of
British Columbia, 1981.

\bibitem{CJP} S. Th\'{e}berge, G. A. Miller and A. W. Thomas, Can. J. Phys.
{\bf 60}, 59 (1982).

\bibitem{DTA} L. R. Dodd, A. W. Thomas, and R. F. Alvarez-Estrada, Phys.
Rev. D {\bf 24}, 1961 (1981); R. F. Alvarez-Estrada and A. W. Thomas, J.
 Phys. G {\bf 9}, 161 (1983).

\bibitem{hedgehog} A. Chodos and C. B. Thorn, Phys. Rev. D {\bf 12}, 2733
(1975); M. C. Birse and M. K. Bannerjee, Phys. Lett. {\bf 136B}, 284 (1984);
Phys. Rev. D {\bf 31}, 118 (1985); S. Kahana, G. Ripka and V. Soni, Nucl.
Phys. {\bf A415}, 351 (1984);
 H. Kitagawa, Nucl. Phys. {\bf A487}, 544  (1988).

\bibitem{gluons} M. Bickeb\"{o}ller, M. C. Birse, and L. Wilets, Z.\
Phys. A {\bf 326}, 89 (1987); M. Bickeb\"{o}ller, M. C. Birse,
M. Marschall and L. Wilets, Phys. Rev. D {\bf 31}, 2892 (1985).

\bibitem{DW} L. R. Dodd and A. G. Williams, Phys. Lett. {\bf 210B}
, 10 (1988).

\bibitem{MBS}J. A. McGovern, M. C. Birse and D. Spanos, J.\ Phys.\ G:
Nucl.\ Part.\ Phys. {\bf 16}, 1561 (1990).

\bibitem{Leech} R. G. Leech and M. C. Birse, J.\ Phys.\ G:
Nucl.\ Part.\ Phys. {\bf 18}, 785 (1992).

\bibitem{Banerjee} M. K. Banerjee, W. Broniowski and T. D. Cohen, in
 {\it Chiral Solitons}, ed. K.-F. Liu (World Scientific, Singapore, 1987),
p. 255.

\bibitem{Bayer} L. Bayer, H. Forkel and W. Weise, Z.\ Phys.\ A {\bf 324}, 365
(1986).

\bibitem{Ban88} M. K. Banerjee, in {\it Quarks, Mesons and Nuclei}, eds.
W.-Y. P. Hwang and E. M. Henley (World Scientific, Singapore, 1988);
C.-Y. Ren and M. K. Banerjee, Phys.\ Rev.\ {\bf C41}, 2370 (1990); M. K.
Banerjee, Phys.\ Rev.\ {\bf C45}, 1359 (1992).

\bibitem{McG} J. A. McGovern, Nucl. Phys. A {\bf 533}, 553 (1991).


\bibitem{Chin}S. A. Chin, Nucl.\ Phys.\ A {\bf382}, 385 (1982).

\bibitem{Aerts}A. Th. M. Aerts, P. J. G. Mulders, and J. J. deSwart,
Phys.\ Rev.\  {\bf D17}, 260 (1978).

\bibitem{Bicke}M. Bickeb\"{o}ller, R. Goldflam and L. Wilets, J.\ Math.\
Phys.\ {\bf 26}, 1810 (1985); P. Tang and L. Wilets, {\it ibid.} {\bf 31}
, 1661 (1990).

\bibitem{CKY}S. A. Chin, A. K. Kerman and X. H. Yang, MIT CTP no. 919,
July 1981; S. A. Chin, Ann.\ of Phys.\ {\bf 108}, 301 (1977).

\bibitem{Det} J.-L. Dethier, R. Goldflam, E. M. Henley and L. Wilets,
Phys.\ Rev.\ {\bf D27}, 291 (1983).

\bibitem{Lub}E. G. Lub\"{e}ck, M. C. Birse, E. M. Henley and L. Wilets,
Phys.\ Rev.\ {\bf D33}, 234 (1986).

\bibitem{DL}L. R. Dodd and M. A. Lohe, Phys.\ Rev.\ {\bf D32}, 1816 (1985).

\bibitem{DWT}L. R. Dodd, A. G. Williams and A. G. Thomas, Phys.
\ Rev.\ {\bf D35}, 1040 (1987).

\bibitem{GDU}R. Goldflam, I. Duck and E. Umland, Phys.\ Lett.\ {\bf 127B},
155 (1983).
\end{references}
\end{document}